\documentclass[submission, Proceedings]{SciPost}

\hypersetup{
    colorlinks,
    linkcolor={red!50!black},
    citecolor={blue!50!black},
    urlcolor={blue!80!black}
}

\usepackage[bitstream-charter]{mathdesign}
\def\a{\alpha}

\def\p{\pi}                     
\def\r{\rho}                    
\def\t{\tau}

\DeclareSymbolFont{usualmathcal}{OMS}{cmsy}{m}{n}
\DeclareSymbolFontAlphabet{\mathcal}{usualmathcal}

\begin{document}

\begin{center}{\Large \textbf{ \boldmath
On the use of the Operator Product Expansion in
finite-energy sum rules for light-quark correlators\\
}}\end{center}

\begin{center}
Diogo Boito,\textsuperscript{1,2}
Maarten Golterman,\textsuperscript{3,4$\star$}
Kim Maltman\textsuperscript{5,6}~and
Santiago Peris,\textsuperscript{4}
\end{center}

\begin{center}
{\bf 1} Instituto de F\'isica de S\~ao Carlos, Universidade de S\~ao Paulo, CP 369, 13560-970, \\ S\~ao Carlos, SP, Brazil
\\
{\bf 2} University of Vienna, Faculty of Physics, Boltzmanngasse 5, A-1090 Wien, Austria
\\
{\bf 3} Department of Physics and Astronomy, San Francisco State University, \\ San Francisco, CA 94132, USA
\\
{\bf 4} Department of Physics and IFAE-BIST, Universitat Aut\`onoma de Barcelona\\
E-08193 Bellaterra, Barcelona, Spain\\
{\bf 5} Department of Mathematics and Statistics, York University,  Toronto, ON Canada M3J~1P3
\\
{\bf 6} CSSM, University of Adelaide, Adelaide, SA~5005 Australia
\\

* maarten@sfsu.edu
\end{center}

\begin{center}
\today
\end{center}

\definecolor{palegray}{gray}{0.95}
\begin{center}
\colorbox{palegray}{
  \begin{minipage}{0.95\textwidth}
    \begin{center}
    {\it  16th International Workshop on Tau Lepton Physics (TAU2021),}\\
    {\it September 27 – October 1, 2021} \\
    \doi{10.21468/SciPostPhysProc.?}\\
    \end{center}
  \end{minipage}
}
\end{center}

\section*{Abstract}
\vspace{-10pt}
{\bf \boldmath
Tau-based finite-energy sum-rule (FESR) analyses often assume that scales
$s_0\sim m_\tau^2$ are large enough that (i) integrated duality violations (DVs)
can be neglected, and (ii) contributions from non-perturbative OPE
condensates of dimension $D$ scale as $(\Lambda_{\rm QCD}/m_\tau )^D$, allowing the
OPE series to be truncated at low dimension. The latter assumption is not
necessarily valid since the OPE series is not convergent, while the former
is open to question given experimental results for the electromagnetic,
$I=1$ vector ($V$), $I=1$ axial vector ($A$) and $I=1$ $V+A$ current spectral functions,
which show DV oscillations with amplitudes comparable in size to the
corresponding $\alpha_s$-dependent perturbative contributions 
at $s\sim2-3$~GeV$^2$. Here, we discuss 
recently introduced new tools for assessing the numerical
relevance of omitted higher-$D$ OPE  contributions.
Applying these to the ``truncated OPE'' strategy used in Refs.~{\cite{Davier:2013sfa,Pich:2016bdg}} and 
earlier work by the same authors, we find that this strategy fails to
yield reliable results for the strong coupling from hadronic $\t$ decays.
}

\noindent\rule{\textwidth}{1pt}\vspace{-10pt}
\tableofcontents\thispagestyle{fancy}\vspace{-2pt}
\noindent\rule{\textwidth}{1pt}

\section{Introduction}
\label{sec:intro}
The determination of the strong coupling $\alpha_s$ from hadronic
$\tau$ decays is interesting for two important reasons:  (i) In principle,
a high precision can be reached, if we ``normalize'' a determination at
the $\tau$ mass scale by evolving the coupling to the $Z$ mass, and
(ii) because of the low scale set by the $\tau$ mass, it provides a 
direct test of the running of the coupling predicted by QCD.
However, at the same time, the strong coupling at the $\tau$ mass,
$\alpha_s(m_\tau)$, is rather large (of order $0.3$ in the $\overline{\rm MS}$
scheme), and non-perturbative effects threaten to contaminate the
extraction of $\alpha_s(m_\tau)$ from experimental data.   It is thus 
important to introduce methods to quantify such contamination, and
to test the resulting strategies for their reliability.   In this talk, we
discuss the ``truncated OPE" (tOPE) strategy \cite{LeDiberder:1992zhd}, which has most 
recently been used in Refs.~\cite{Davier:2013sfa,Pich:2016bdg}.
We will demonstrate that this strategy leads to  results
with unquantifiable systematic errors at the scale of the current desired level of precision, and hence should no longer be used.  

The determination of $\alpha_s(m_\tau)$ starts by considering 
finite-energy sum rules (FESRs) of the form\footnote{For a more
detailed description of our use of FESRs, see Ref.~\cite{Boito:2016oam}
and references therein.}
\begin{equation}
\label{FESR}
\int_0^\infty ds\,s^n\,\rho(s)=-\frac{1}{2\pi i}\oint_{|z|=s_0} dz\,z^n\,\Pi(z)\ ,
\end{equation}
where $\Pi(z)$ is the (scalar) vacuum polarization obtained from 
the $V+A$ non-strange $I=1$ or the electro-magnetic (EM) spectral function $\rho(s)$ (with $s=q^2>0$), and the contour $|z|=s_0$ is a circle in the  complex $q^2=z$ plane around the origin with radius $s_0$,
which, if $\rho(s)$ is obtained from hadronic $\tau$ decays, is bounded
from above by the $\tau$ mass, $s_0\le m_\tau^2$.

Equation~(\ref{FESR}) is exact.   To proceed, we then approximate
\begin{equation}
\label{Pi}
\Pi(z) = \Pi_{\rm pert.}(z)+\Pi_{\rm OPE}(z)+\Pi_{\rm DV}(z)\ ,
\end{equation}
where $\Pi_{\rm pert.}(z)$ is the perturbative part, for which a 
five-loop expression exists \cite{Baikov:2008jh}, $\Pi_{\rm OPE}(z)$
is the operator-product-expansion (OPE) part,\footnote{For the non-strange channel, the $D=2$ term proportional to the square of the
light quark masses can be safely neglected.   We have also checked that 
the logarithmic dependence of the $C_D$ on $q^2$ can be neglected
\cite{Boito:2011qt}.}
\begin{equation}
\label{PiOPE}
\Pi_{\rm OPE}(q^2)=\frac{C_4}{(q^2)^2}-\frac{C_6}{(q^2)^3}+\frac{C_8}{(q^2)^4}+\dots\ ,
\end{equation}
and $\Pi_{\rm DV}(z)$ is the part violating quark-hadron duality, which
is not captured by the OPE \cite{Boito:2017cnp}.

The existence of a duality-violating (DV) part is closely related to the fact that the OPE
is not convergent; rather it is (at best) an asymptotic expansion.   
According to expectations, the DV part, which physically represents the oscillations
about perturbative
expectations associated with 
the presence of resonances visible in the spectral function, decreases exponentially
with $q^2$, {\it i.e.}, non-perturbatively in the OPE expansion parameter
$1/q^2$.    We note that the weight $z^n$ in Eq.~(\ref{FESR}) picks out the term proportional to $1/q^{2(n+1)}$ in the OPE---this will be important
in what follows.

\begin{figure}[t]
\centering
\includegraphics[width=0.6\textwidth]{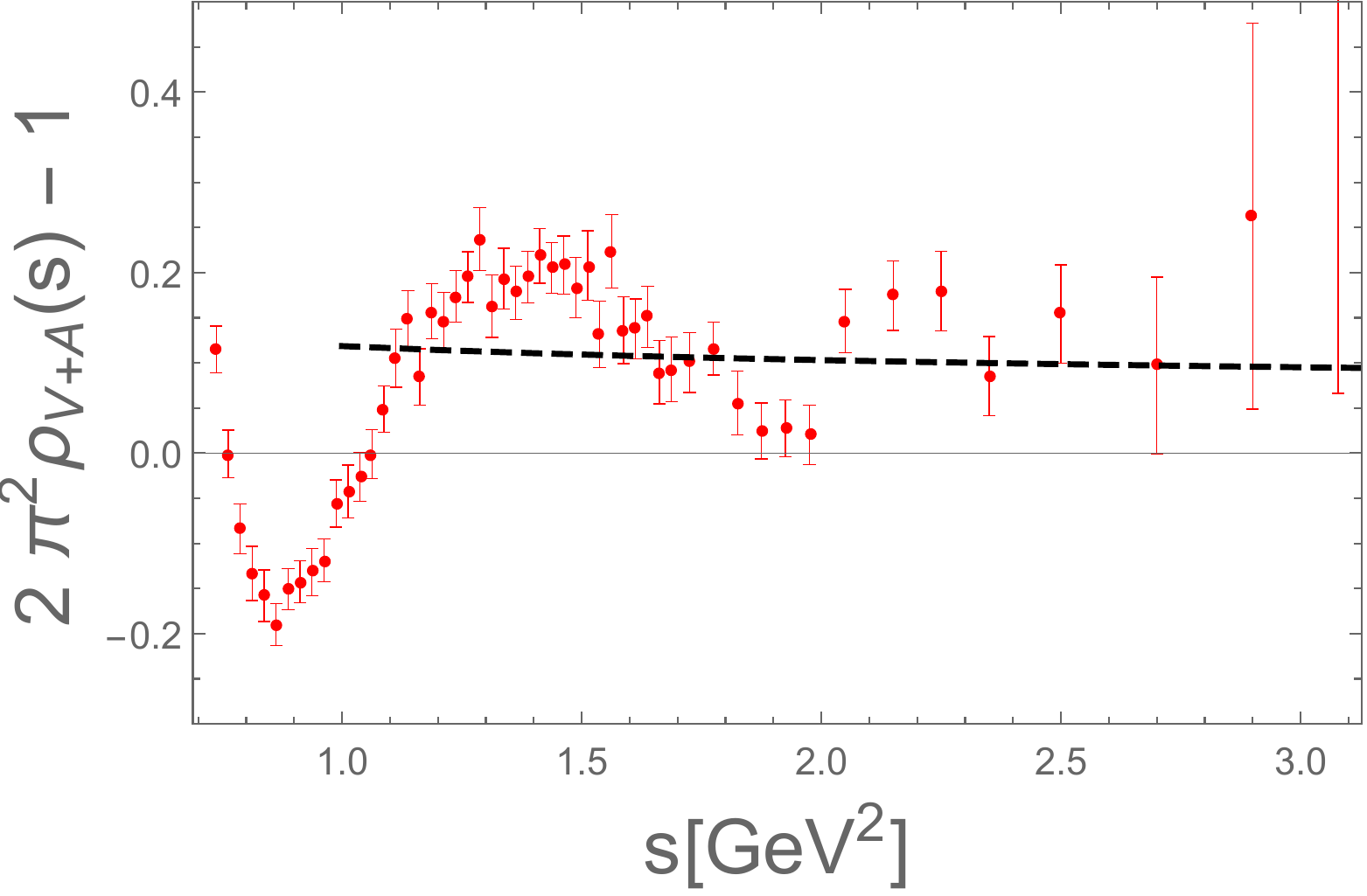}
\caption{Blow-up of the ALEPH data (red data points) \cite{Davier:2013sfa} in the
large-$s$ region of the $V+A$ non-strange spectral function. What is shown is $2\p^2\r_{V+A}-1$, {\it i.e.},
the dynamical QCD contribution to the spectral distribution.
Black dashed line:  perturbation theory, also with the parton-model contribution subtracted.}
\label{fig:spec}
\end{figure}

While the DV part is exponentially suppressed, the data are limited to
$s\le m_\t^2$, and we are forced to consider the possibility that they may not be
negligible.   In this respect, it is instructive to consider Fig.~\ref{fig:spec}.
The oscillations in the spectral function (red data points) represent the presence of resonances, and are not captured by the OPE (black dashed curve).
It is clear that DVs, which represent the oscillations of the data around the dashed curve, are not a small part of the dynamical QCD contribution to the spectral function in this region.  Note that in order to make this comparison, one should subtract the parton-model contribution to the spectral function, as it is independent of
$\alpha_s$ and thus not part of 
the dynamics produced
by QCD.

\section{The truncated-OPE strategy}
\label{sec:tOPE}
Two different strategies have been developed to deal with the non-perturbative contamination, {\it i.e.}, the $D>0$ terms in the OPE and DVs:  the tOPE strategy, and the ``DV-model'' strategy.\footnote{A variant of the tOPE strategy can be found in Ref.~\cite{Ayala:2021mwc}.}  For the latter, 
we refer to Boito's talk at this workshop \cite{Boitotalk}; the most recent application of the DV-model strategy can be found in Ref.~\cite{Boito:2020xli}.

The assumptions underlying the tOPE strategy are the following. 
First, DVs are neglected, but the dangerous region, where the 
circular contour on the right-hand side of Eq.~(\ref{FESR}) crosses
the positive real axis, is suppressed by combining the weights of
Eq.~(\ref{FESR}) into polynomials with multiple zeroes at $s=s_0$.
These multiple zeroes are thus introduced to suppress DVs, at the intersection of the
contour on the right-hand side of
Eq.~(\ref{FESR}) with the positive real axis.
Furthermore, $s_0$ is typically chosen equal to $m_\t^2$, in order
to keep values of $\a_s(s_0)$ appearing in the fit as small as 
possible.

In implementations of the tOPE strategy in the literature, these polynomial have degrees varying between 3 and 7.  This 
implies that OPE terms up to order $D=16$ contribute to the 
right-hand side of  Eq.~(\ref{FESR}), and one thus would have
to fit $\a_s(m_\t)$ as well as $C_{4,6,8,10,12,14,16}$, eight parameters
in total.   However, the number of independent polynomials of maximal degree 7 with at least two zeroes
at $s=s_0$ is smaller than eight, and thus, necessarily, a number
of parameters have to be set equal to zero by hand.

One such set of polynomials,
used in Ref.~\cite{Pich:2016bdg} and referred to as "optimal" there, is the set
\begin{eqnarray}
\label{optimal}
w_{2n}=1-(n+2)x^{n+1}+(n+1)x^{n+2}\ ,\quad n=1\ ,\dots,\ 5\ ,
\end{eqnarray}
with $x=s/s_0$.   These weights have a double zero at $s=s_0$
(they are ``doubly pinched''), and probe $\a_s(s_0)$ as well as
$C_{6,8,10,12,14,16}$.   With five weights, and taking $s_0=m_\t^2$,
one has five data points, which leads to the choice to set $C_{12}=C_{14}=C_{16}=0$, so that one has four parameters for a fit to five data points,\footnote{The weights of Eq.~(\ref{optimal}) do not project on the $D=4$ term in the OPE.}   Different, but similar, sets of weights have been considered
in Ref.~\cite{Pich:2016bdg} as well as in our analysis of this strategy
\cite{Boito:2016oam}.   Here we will only discuss the set (\ref{optimal}),
as the conclusions from our more extensive study of other sets of weights is the same.   Finally, in the tOPE strategy, one primarily  
considers the $V+A$ channel, as DV and OPE-truncation effects are
argued to be less severe in $V+A$ than in the $V$ or $A$ channels
separately.

In a first look at the tOPE strategy, let us compare two choices for the
OPE coefficients $C_{12}$, $C_{14}$, and $C_{16}$, which are not part of the
fit, and for which thus {\it a priori} values have to be chosen.   We
compare the choice of Refs.~\cite{Davier:2013sfa,Pich:2016bdg},
which effectively sets $C_{12}=C_{14}=C_{16}=0$ (choice 1), with
choice 2:
\begin{equation}
\label{choice2}
C_{12}=0.161~\mbox{GeV}^{12}\ ,\quad 
C_{14}=-0.17~\mbox{GeV}^{14}\ ,\quad
C_{16}=-0.55~\mbox{GeV}^{16}\ .
\end{equation}
This choice is equally arbitrary, but equally reasonable.   
The results of applying the tOPE strategy with either of these two
choices are:
\begin{equation}
\label{table}
\begin{array}{|c|c|c|c|c|c|}
\hline
& \a_s(m_\t) & C_6~(\mbox{GeV}^6) & C_8~(\mbox{GeV}^8) & C_{10}~(\mbox{GeV}^{10}) & \chi^2/\mbox{dof} \\
\hline
\mbox{choice}~1 & 0.317(3) & 0.0014(4) & -0.0010(5) & 0.0004(3) & 1.26/4\\
\hline
\mbox{choice}~2 & 0.295(4) & -0.0130(4) & 0.0356(5) & -0.0836(3) & 1.09/1 \\
\hline
\end{array}
\end{equation}
The fits were done with FOPT \cite{Boito:2016oam}, and only the 
fit errors are shown.   The OPE coefficients in Eqs.~(\ref{choice2})
and (\ref{table}) are very reasonable, increasing in absolute value 
with the order in the OPE, but consistent with it being an asymptotic
expansion.  

Clearly, the choice-1 and choice-2 fits are inconsistent, and lead to 
values of $\a_s(m_\t)$ which are about 7\% apart; this
is about double the total error quoted in Ref.~\cite{Pich:2016bdg}.
Moreover, there is no way to tell which of these two fits is closer to
the truth; in fact, both fits may be wrong.

\section{Tests of the truncated-OPE strategy on data for $e^+e^-\to$~hadrons}
\label{sec:EM}

In order to probe this unsatisfactory state of affairs in more detail,
we will apply the tOPE strategy next to $R$-ratio data obtained from
$e^+e^-\to$~hadrons.    The key observation is, of course, that if the
tOPE strategy works at $s_0=m_\t^2$, as necessitated with data
from $\t$ decays, it should certainly work at $s_0>m_\t^2$, where
$R$-ratio data are available.

There are, of course, differences with the $\t$-based approach.
First, $e^+e^-\to$~hadrons only gives access to $V$ channel data,
whereas it is advocated to apply the tOPE strategy to $V+A$.
However, Refs.~\cite{Davier:2013sfa,Pich:2016bdg} find that 
the tOPE strategy applied to the $V$ channel $\t$-decay data 
yields results consistent with those from $V+A$, with 
equally good fit qualities.    Another difference is that $R$-ratio
data contain an $I=0$ component, in addition to the $I=1$ 
component related to the $\t$-based $V$ spectral function.  However,
the $I=0$ component is an $SU(3)$-flavor partner of the $I=1$
component, and it is known that the strange quark mass that
breaks $SU(3)$ has a very small effect on $\a_s$ \cite{Boito:2018yvl}.
We thus conclude that lessons learned from applying the 
tOPE strategy to $R$-ratio data are relevant for the application
to $\t$-based analyses at as well.   Below we will use $R$-ratio data from 
Refs.~\cite{Keshavarzi:2018mgv,Keshavarzi:2019abf}.

First, we repeat the fit with weights (\ref{optimal}) at $s_0\approx m_\t^2$, now using the
$R$-ratio data.   We find\footnote{We omit
results for the OPE coefficients $C_{D>0}$ for this discussion.} 
\begin{eqnarray}
\label{firstR}
&\chi^2~\mbox{fit}: & \a_s(m_\t)=0.308(4)\ ,\qquad p\mbox{-value}=2\times 10^{-15}\ ,\\
&\mbox{diagonal~fit}: & \a_s(m_\t)=0.245(10)\ .\nonumber
\end{eqnarray}
This is clearly a disaster.   We note that a diagonal fit uses a diagonal 
fit quality, but takes the full data covariance matrix into account for
error propagation; for more detail on this, see Refs.~\cite{Boito:2011qt,Boito:2019iwh}.
Next, let us take $s_0$ larger: if we take $s_0=3.6$~GeV$^2$, the
$p$-value of the $\chi^2$ fit becomes larger than 10\%.  In fact, at $s_0=3.6$~GeV$^2$, we find
\begin{eqnarray}
\label{secondR}
&\chi^2~\mbox{fit}: & \a_s(m_\t)=0.264(5)\ ,\qquad p\mbox{-value}=0.41\ ,\\
&\mbox{diagonal~fit}: & \a_s(m_\t)=0.256(12)\ .\nonumber
\end{eqnarray}
This is a clear improvement, but it yields a very low value for $\a_s$:
at the $Z$ mass, this would translate into $\a_s(m_Z)=0.110$!
Other sets of weights considered in Refs.~\cite{Davier:2013sfa,Pich:2016bdg} lead to very similar results.   For more details, we refer to
Ref.~\cite{Boito:2019iwh}.   

One might argue that the tests of Eqs.~(\ref{firstR},\ref{secondR}) are
possibly somewhat inconclusive.  
However, the $R$-ratio data allow us to subject the tOPE strategy to a 
more stringent test, by considering the $s_0$ dependence of tOPE-based
fits.   Again, the simple observation is that if the tOPE strategy works
at values of $s_0\ge m_\t^2$, there should be a good match between 
theory and experiment for {\em all} values of $s_0\ge m_\t^2$.

\begin{figure}[h!]
\centering
\includegraphics[width=0.45\textwidth]{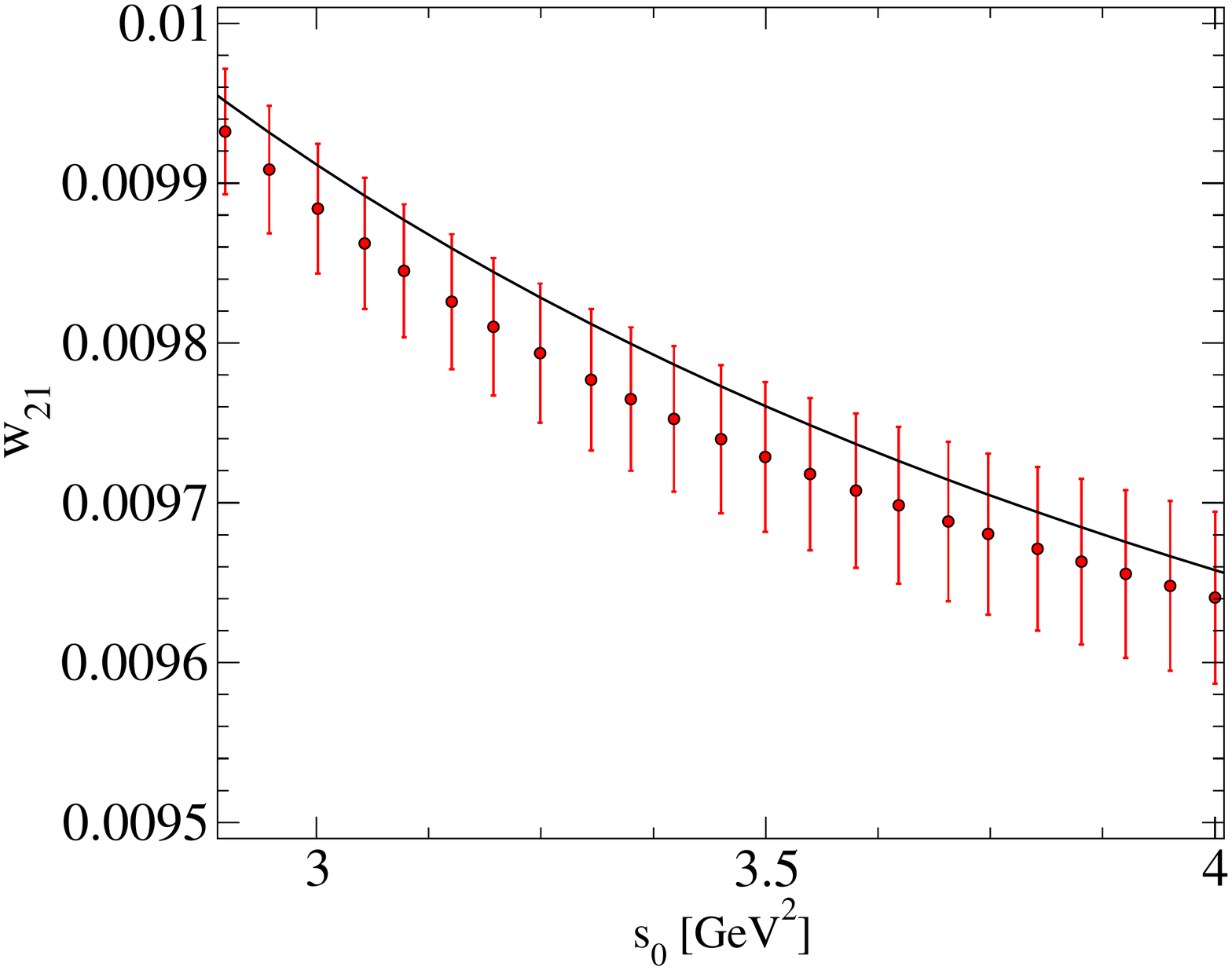}
\includegraphics[width=0.45\textwidth]{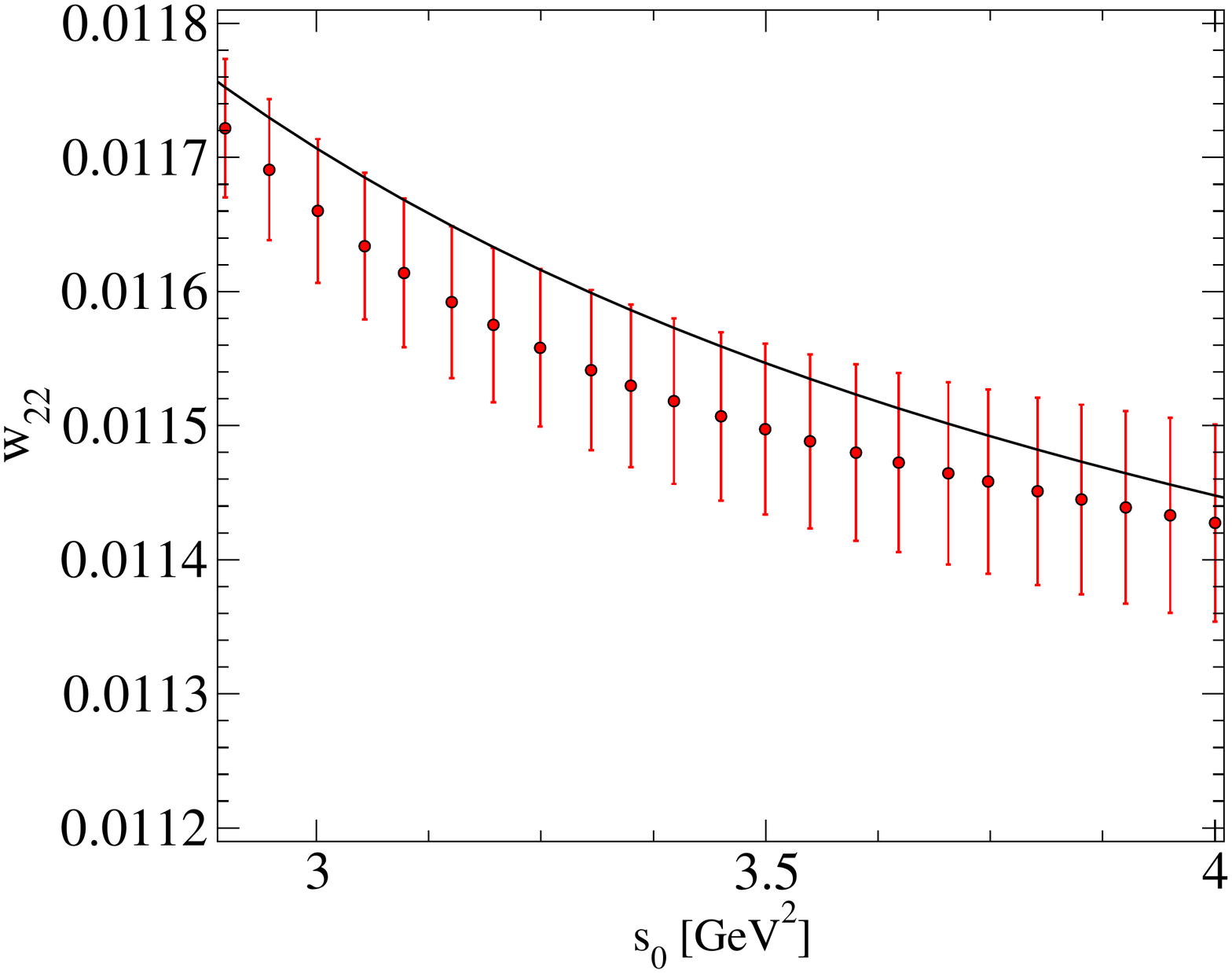}
\includegraphics[width=0.45\textwidth]{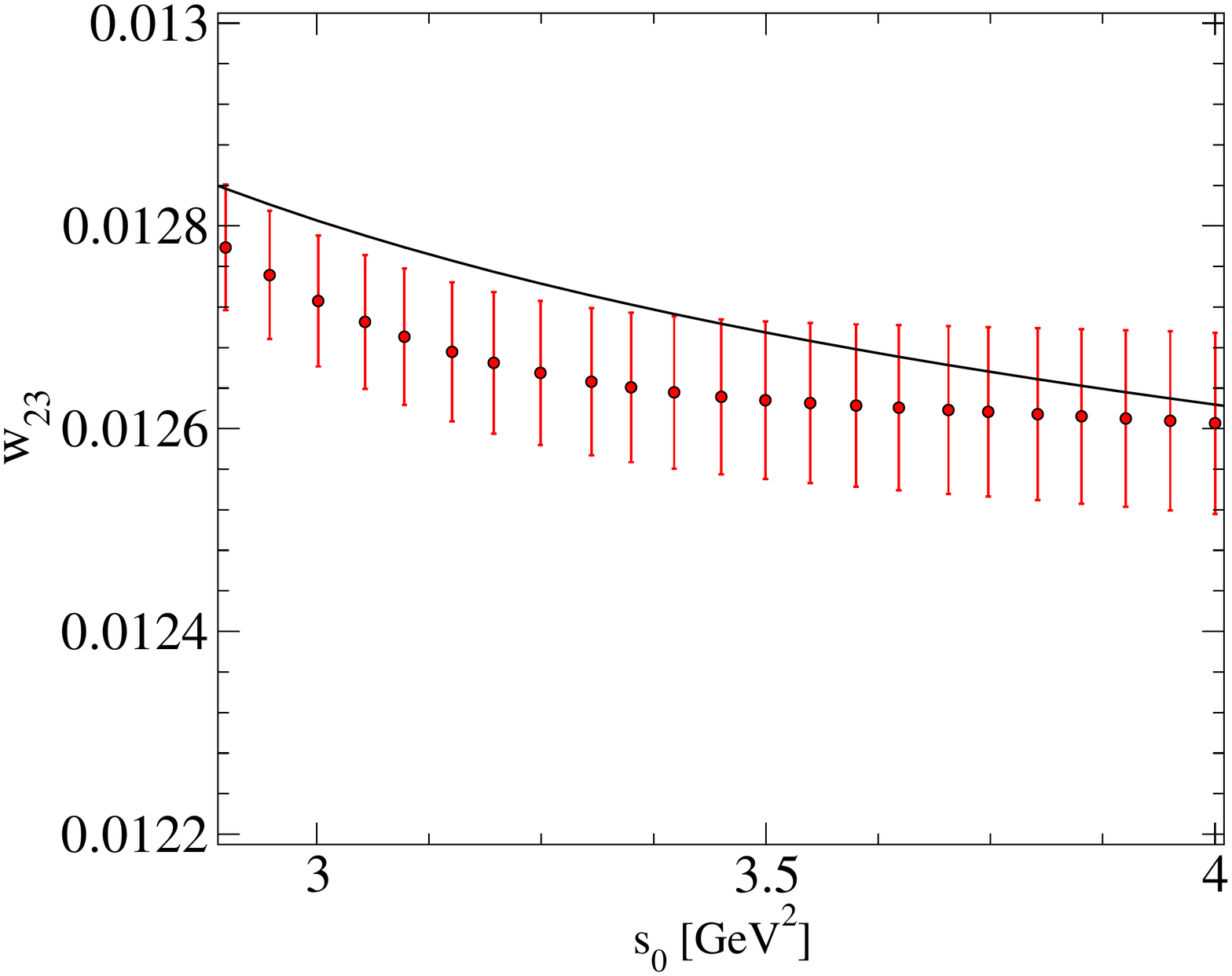}
\includegraphics[width=0.45\textwidth]{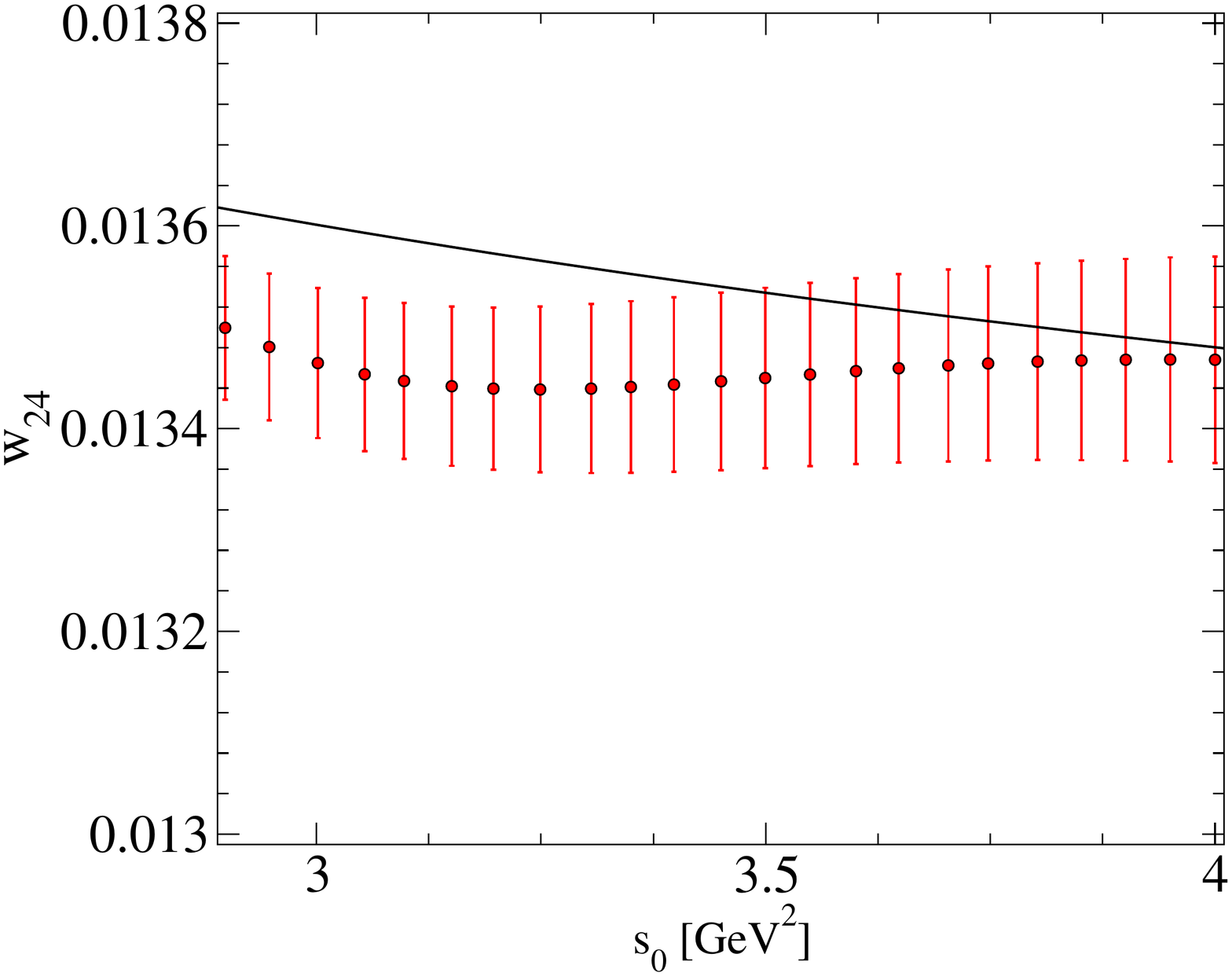}
\includegraphics[width=0.45\textwidth]{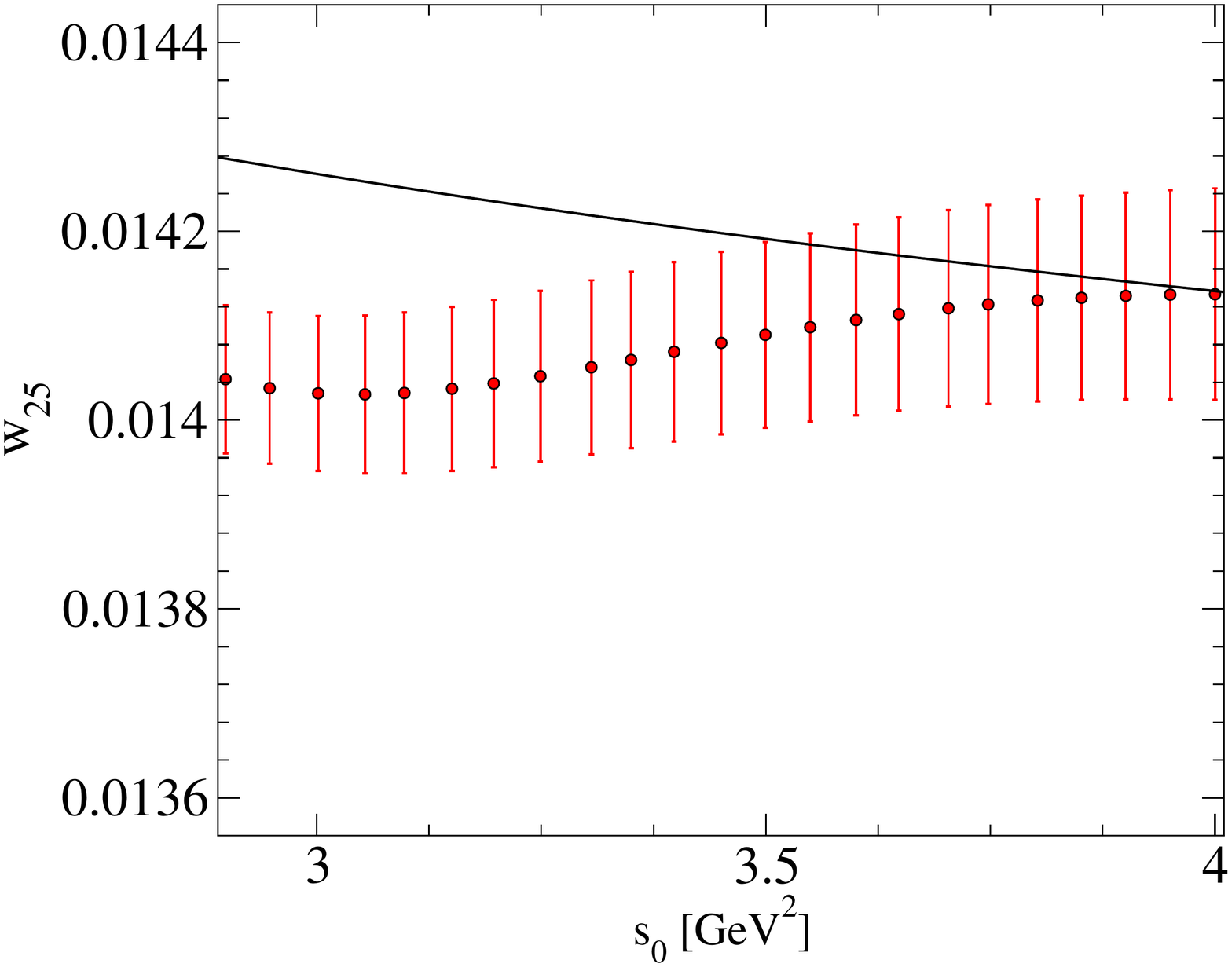}
\caption{tOPE fits using optimal weights with $s_0=s_0^*=3.6$~GeV$^2$
in Eq.~(\ref{FESR}).  The red data points show the
left-hand side of Eq.~(\ref{FESR}) for each weight; the black curves 
show the fits of the right-hand side.}
\label{fig:match}
\end{figure}
Figure~\ref{fig:match} shows the left-hand side and the right-hand side
of Eq.~(\ref{FESR}) obtained in 
tOPE $\chi^2$ fits employing optimal weights at
a fixed $s_0\equiv s_0^*=3.6$~GeV$^2$, as a function of $s_0$.   While superficially,
one might conclude that the agreement between experiment
(left-hand side of Eq.~(\ref{FESR})) and theory (right-hand side Eq.~(\ref{FESR})) as a function of $s_0$ is not unreasonable, this is not actually the case.   In fact, it is not easy to judge the level of agreement
visually, as there are strong correlations, both between the spectral integrals at different $s_0$, between the fitted theory integrals at different $s_0$, and between the theory integrals and the spectral integrals used to fit the parameters of the theory representation.   A careful look reveals a
possible sign of trouble: clearly the slopes of the fit curves at $s_0>s_0^*$
are rather different than the corresponding slopes in the data, for the
weights $w_{23}$, $w_{24}$ and $w_{25}$.

Whether this difference in slopes is statistically significant or not can be
investigated by considering the double differences
\begin{equation}
\label{doublediffs}
\Delta^{(2)}(s_0;s_0^*)=\left[I^{\rm th}_w(s_0)-I^{\rm exp}_w(s_0)\right]-
\left[I^{\rm th}_w(s_0^*)-I^{\rm exp}_w(s_0^*)\right]\ ,
\end{equation}
where $I^{\rm exp}_w(s_0)$ denotes the left-hand side of Eq.~(\ref{FESR}) for polynomial weight $w$, and 
$I^{\rm th}_w(s_0)$ denotes the right-hand side of Eq.~(\ref{FESR}).
These double differences compare theory ({\it i.e.}, the fits) with
experiment, relative to a reference value $s_0^*$.   Of course, it is 
important, in computing these double differences, to take all 
correlations, including those between data and fitted parameters,
into account.   These double differences should be consistent with
zero for the tOPE strategy to pass this type of test.
\begin{figure}[th!]
\centering
\includegraphics[width=0.45\textwidth]{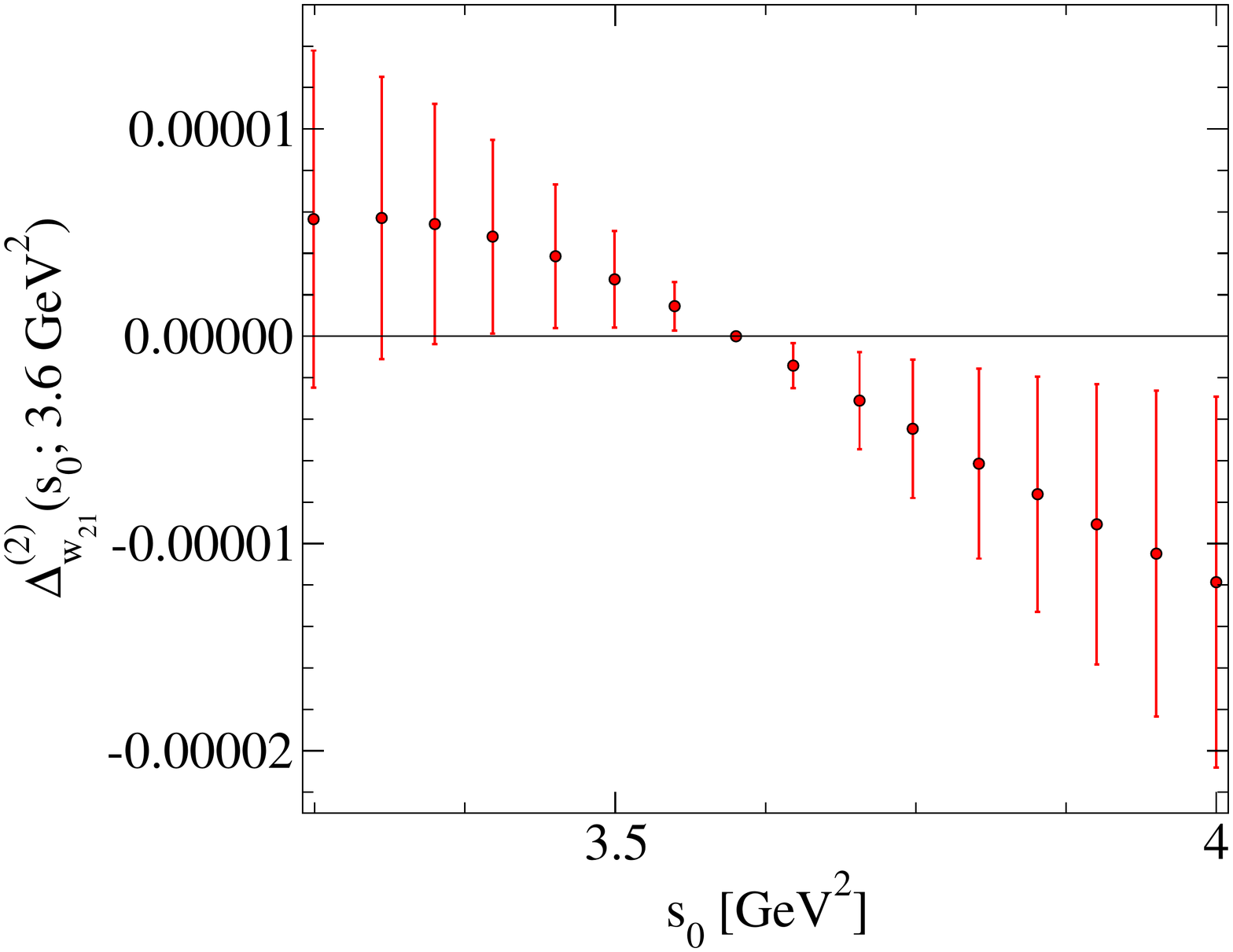}
\includegraphics[width=0.45\textwidth]{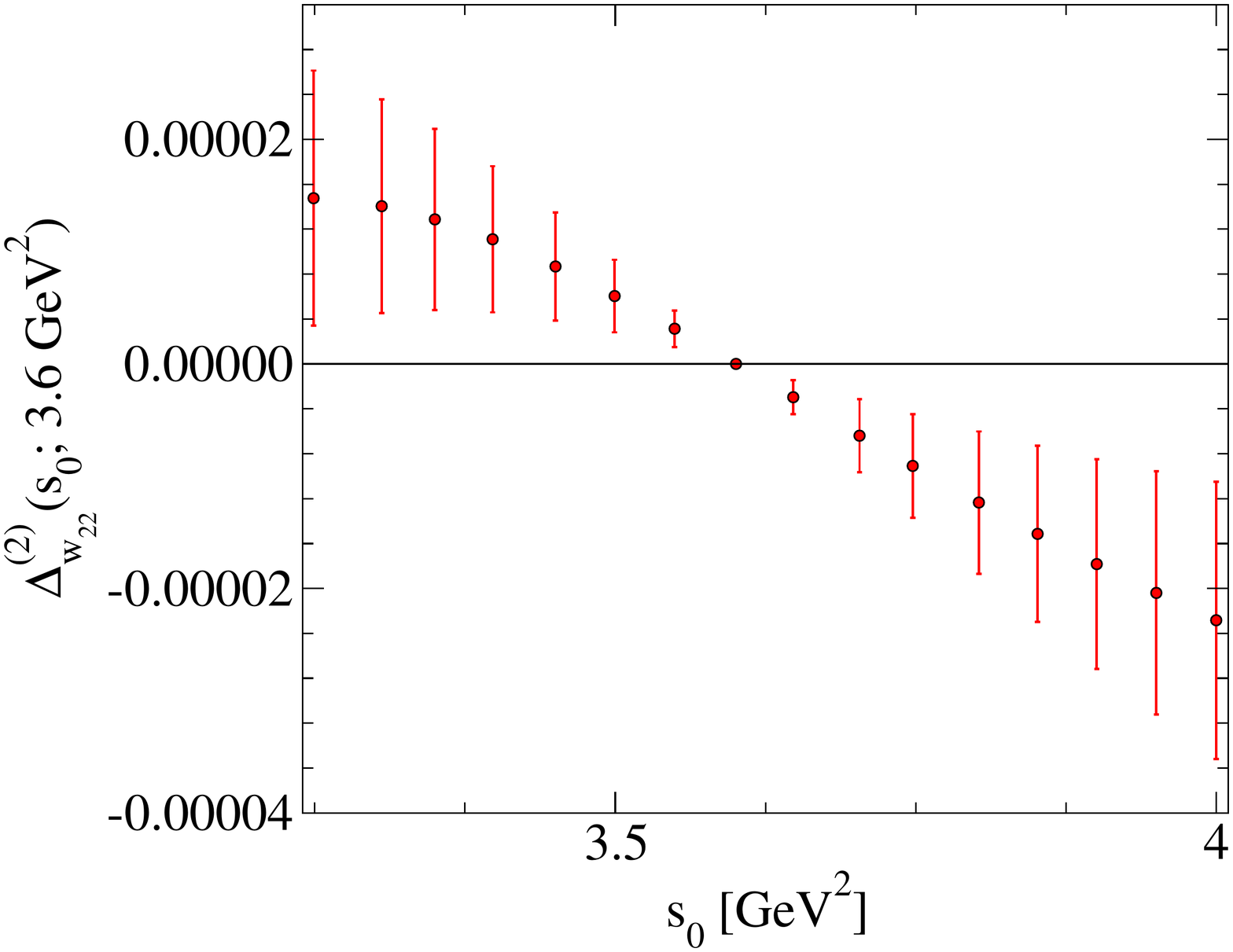}
\includegraphics[width=0.45\textwidth]{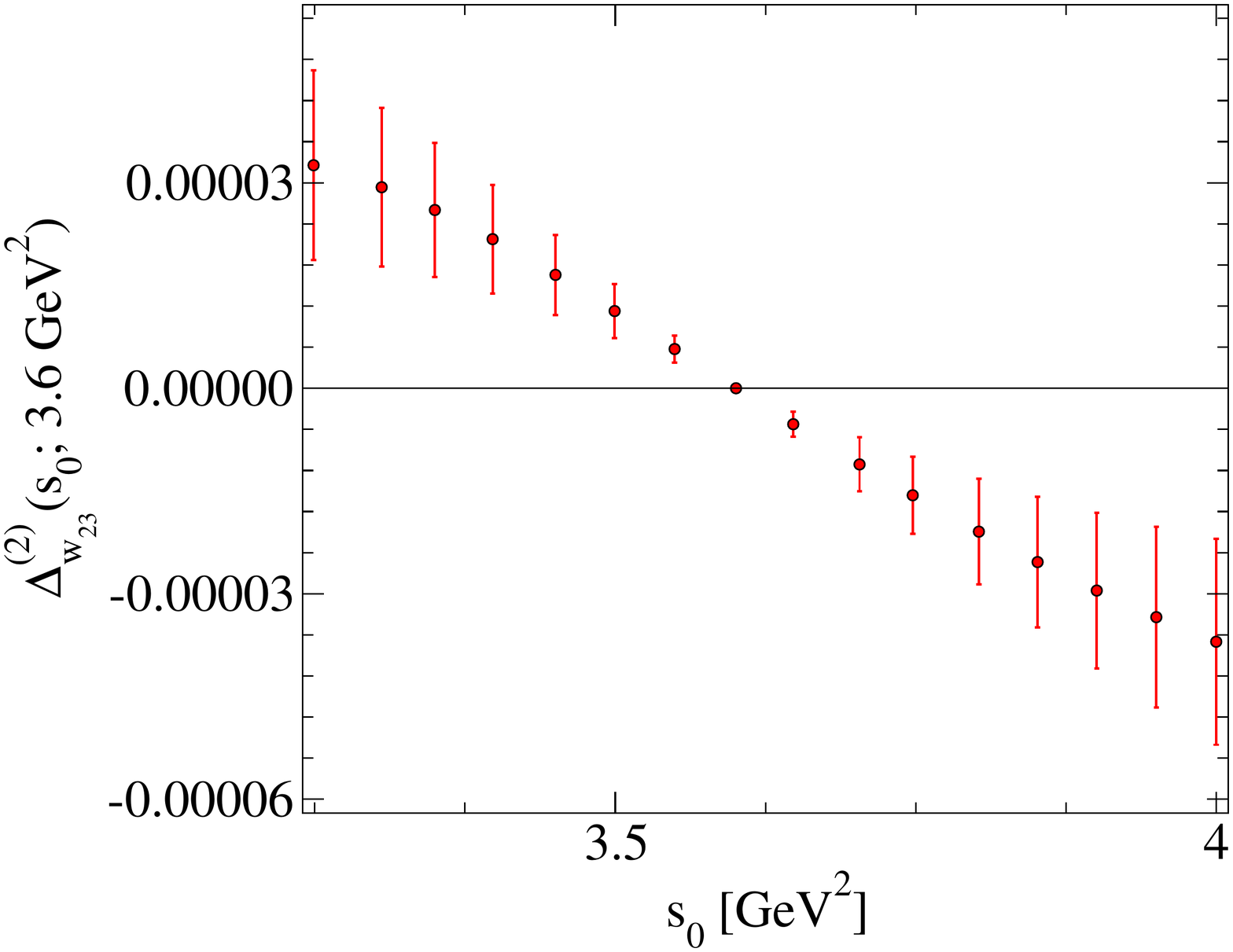}
\includegraphics[width=0.45\textwidth]{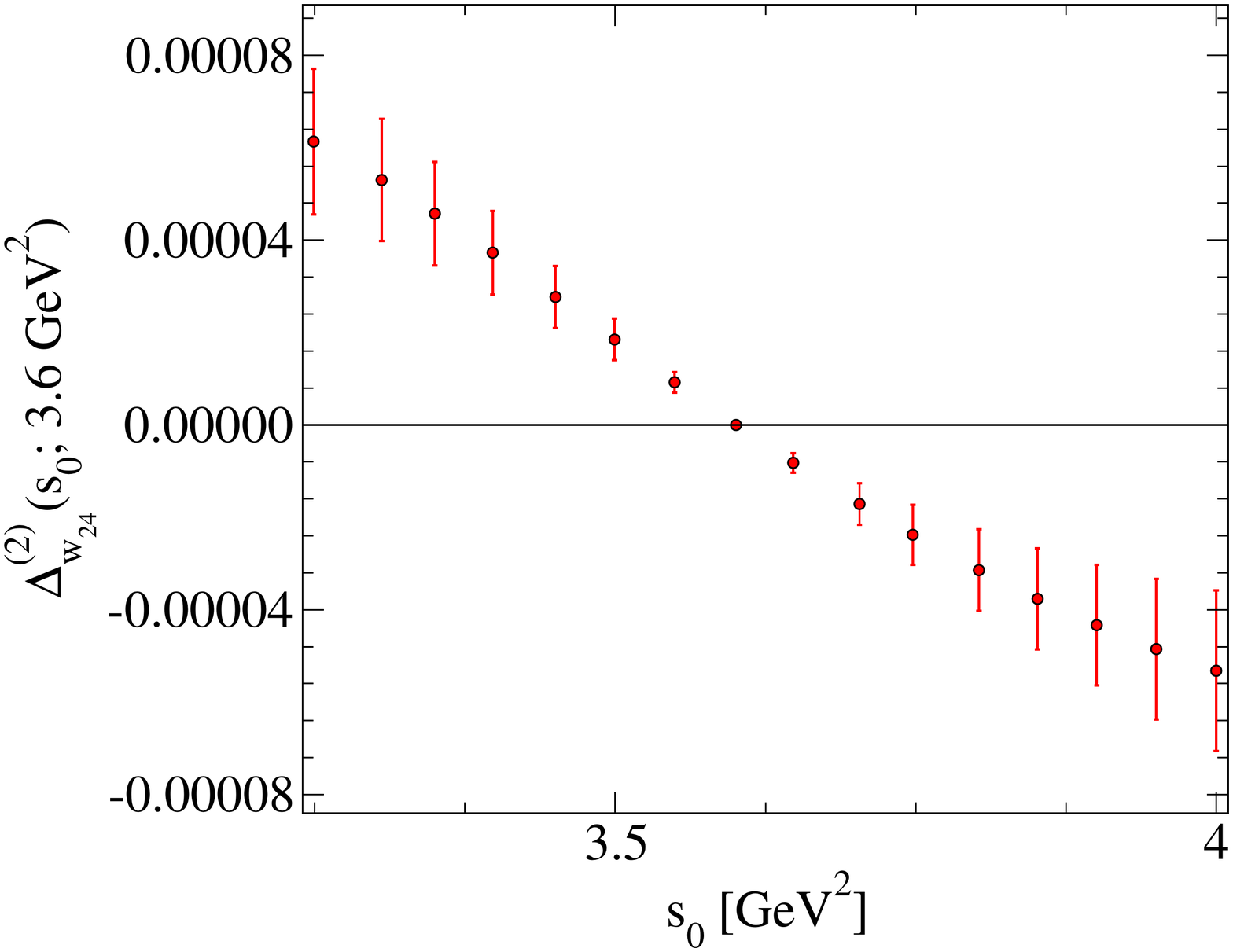}
\includegraphics[width=0.45\textwidth]{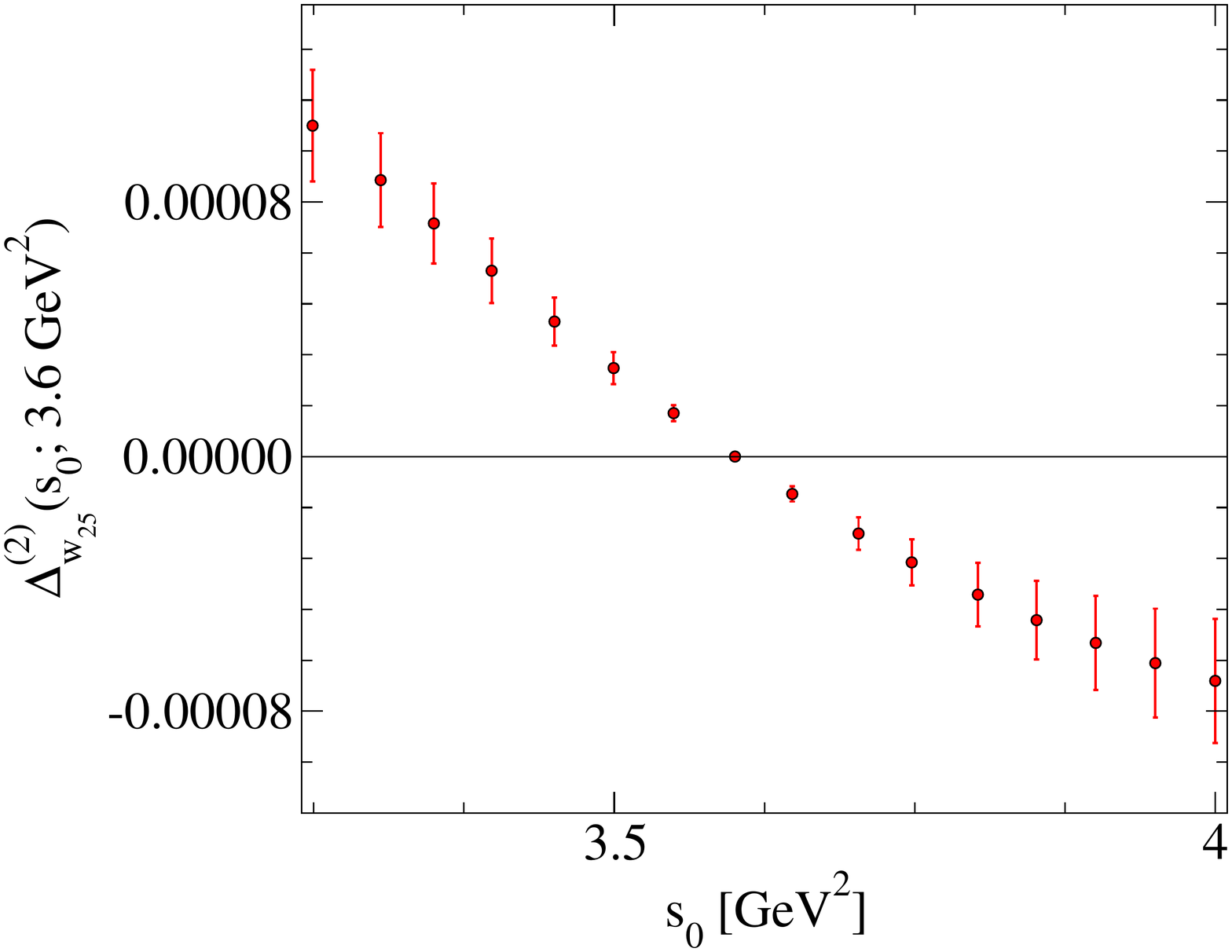}
\caption{The double differences $\Delta^{(2)}(s_0;s_0^*)$ as a function of
$s_0$, for $s_0^*=3.6$~GeV$^2$.  See text.}
\label{fig:diffs}
\end{figure}

Figure~\ref{fig:diffs} shows the double differences for the fits shown in
Fig.~\ref{fig:match}.   Clearly, for at least the weights $w_{23}$, $w_{24}$
and $w_{25}$, they are very far from consistent with zero, for values
of $s_0$ on both sides of $s_0^*$.  (By construction, the double 
differences vanish exactly at $s_0=s_0^*$.)   From these
(and other, see Ref.~\cite{Boito:2019iwh}) $R$-ratio based tests,
we conclude that the tOPE strategy fails.   Because of the close
similarity between the EM and $\t$-based spectral functions, it is
clear that also for hadronic $\t$ decays, the tOPE cannot be trusted
to yield reliable results.

\section{Conclusion}
Since the $\t$ mass is relatively light, one has  deal with the question of possible
non-perturbative contamination in any strategy to determine
the strong coupling from hadronic $\t$-decay data.   In order to 
do this, assumptions are needed, and these assumptions need to
be tested.   Here we considered the truncated-OPE strategy, in
which the main assumption is that higher-order terms in the OPE
can be neglected, and thus effectively be set equal to zero.   This
constitutes an arbitrary choice,
and it is particularly dangerous, given the asymptotic nature of the OPE, when working with FESRs involving weights for which unsuppressed contributions of high dimension
are in principle present.   We carried out 
several tests of the tOPE strategy, quantitatively probing the validity 
of the assumption made about the OPE.

We found that indeed this assumption cannot be trusted at scales of order the $\t$ mass:   The
tOPE strategy does not pass EM-based self-consistency tests, 
described in Sec.~\ref{sec:EM}.  Moreover, in Sec.~\ref{sec:tOPE}
we also showed that it does not pass $V+A$ $\t$-based 
self-consistency tests.   Our conclusion is that the tOPE strategy
is not reliable if the goal is to obtain $\a_s(m_\t)$ with currently competitive 
accuracy.   Values of $\a_s(m_\t)$ obtained with this approach
depend very strongly on arbitrary assumptions made about the 
OPE; these assumptions are not based on QCD.  

For a much more detailed description and discussion of this work,
as well as many more references, we refer to Refs.~\cite{Boito:2016oam,Boito:2019iwh}.

\section*{Acknowledgements}
We would like to thank the organizers of the Tau2021 workshop for organizing this conference series during difficult times.  We also
thank Alex Keshavarzi, Daisuke Nomura and Thomas Teubner for
making their compilation of $R$-ratio data available to us.

\paragraph{Funding information}
DB was supported by the S\~ao Paulo Research Foundation (FAPESP)
Grant No.~2015/20689-9, by CNPq Grant No.~309847/2018-4, and by Coordena\c c\~ao de Aperfei\c coamento de Pessoal de N\'ivel Superior -- Brasil (CAPES) -- Finance Code 001.
MG is supported by the U.S.\ Department of Energy,
Office of Science, Office of High Energy Physics, under Award No.
DE-SC0013682.
KM is supported by a grant from the Natural Sciences and Engineering
Research Council of Canada.
SP is supported by CICYTFEDER-FPA2017-86989-P and by Grant No. 2017 SGR 1069.
IFAE is partially funded by the CERCA program of the Generalitat de Catalunya.

\bibliography{References}

\nolinenumbers

\end{document}